\documentstyle[12pt]{article}
\newcommand{\reals}{\mbox{${\rm I\!R }$}}

\newcommand{\ant}{\int\limits}

\newcommand{\sleu}{\sum_{l=1}^{\infty}}

\newcommand{\cam}{{\cal M}}
\newcommand{\caz}{{\cal Z}}

\newcommand{\beq}{\begin{eqnarray}}
\newcommand{\eeq}{\end{eqnarray}}
\newcommand{\nn}{\nonumber}

\newcommand{\back}{\bar{\varphi}}
\newcommand{\abl}{\partial}

\newcommand{\cq}{\left(\frac{ML}{2\pi}\right)}

\def\bce{\begin{center}}
\def\ece{\end{center}}
\def\bea{\begin{eqnarray}}
\def\eea{\end{eqnarray}}
\def\brr{\begin{array}}
\def\err{\end{array}}
\def\ben{\begin{enumerate}}
\def\een{\end{enumerate}}
\def\bei{\begin{itemize}}
\def\eei{\end{itemize}}

\def\ms{\medskip}

\begin{document}

\hfill UB-ECM-PF 94/3

\hfill February 1994

\vspace*{3mm}

\begin{center}

{\LARGE \bf
One-loop effective potential in $2D$ dilaton gravity on hyperbolic
plane}

\vspace{4mm}

\renewcommand
\baselinestretch{0.8}
\ms

\renewcommand
\baselinestretch{1.4}
{\sc K. Kirsten}\footnote{E-mail: klaus@ebubecm1.bitnet. Alexander von
Humboldt Foundation Fellow.}
and {\sc S.D. Odintsov}\footnote{E-mail: odintsov@ebubecm1.bitnet.
On leave from: Tomsk Pedagogical Institute, 634041 Tomsk, Russian
Federation.} \\  Department E.C.M., Faculty of Physics,
University of  Barcelona, \\  Diagonal 647, 08028 Barcelona,
Catalonia,
Spain \\

\renewcommand
\baselinestretch{1.4}

\vspace{5mm}

{\bf Abstract}

\end{center}
The one-loop effective potential in $2D$ dilaton gravity in conformal
gauge on the topologically non-trivial plane $\reals \times S^1$ and on
the hyperbolic plane $H^2/\Gamma$ is calculated. For arbitrary choice
of
the tree scalar potential it is shown, that the one-loop effective
potential
explicitly depends on the reference metric (through the dependence on
the radius of the torus or the radius of $H^2/\Gamma$). This phenomenon
is absent only for some special choice of the tree scalar potential
corresponding to the Liouville potential and leading to one-loop
ultraviolet
finite theory. The effective equations are discussed and some
interpretation of the reference metric dependence of the
effective potential is made.

\vspace{4mm}


\newpage
There was recently much activity in the study of $2D$ dilaton gravity
which was considered as a toy model for a still unknown consistent
theory of $4D$ quantum gravity. One of the main motivations for such a
study was the identification of black holes in such a theory (see,
original works \cite{callangiddingsharveystrominger92,witten91}).

Different theories of $2D$ dilaton gravity (mainly string-inspired
models)
have been discussed (see \cite{harvey} for a review). The model of $2D$
dilaton gravity which is quite general and which will be the model we
are interested in is described by the following action
\cite{odintsovshapiro91}
\beq
S=\int d^2x|g|^{\frac 1 2}\,\,\left[\frac 1 2 g^{\mu\nu}\abl
_{\mu}\varphi \abl_{\nu}\varphi +c \varphi R +V(\varphi
)\right],\label{1}
\eeq
where $\varphi$ is a dimensionless scalar, $c$ is the scalar-graviton
coupling constant and $V(\varphi )$ is the scalar potential. Note that
the action (\ref{1}) can be transformed on the classical level
\cite{russotseytlin92} to the form corresponding to the string effective
action (for an introduction to string theory, see book
\cite{greenschwarzwitten87}).

The renormalization properties of the theory (\ref{1}) are known, in
particularly the direct calculations in a
general covariant gauge \cite{odintsovshapiro91} or in
conformal gauge \cite{russotseytlin92} show that the $2D$ dilaton
gravity (\ref{1}) is multiplicatively renormalizable in the usual sense
for the potential chosen in Liouville form:
\beq
V(\varphi )= \mu^2 e^{\alpha \varphi },\label{2}
\eeq
where $\alpha$ is an arbitrary parameter.
In this case, the action (\ref{1}) corresponds to the celebrated
Liouville theory.

Recently, the study of the off-shell one-loop effective potential in the
theory (\ref{1}) in the general covariant gauge
\cite{odintsovshapiro91} has been performed on flat
background \cite{baninshapiro93}. As it usually happens in gauge
theories, is was found \cite{baninshapiro93} that this effective
potential is highly gauge dependent (through the explicit dependence
from the gauge parameters).

In ref.~\cite{russotseytlin92} the calculation of the one-loop effective
potential was performed in the conformal gauge
\beq
g_{\mu\nu} = e^{2\sigma }\eta _{\mu\nu} \label{3}
\eeq
on flat background with constant $\varphi$. In the conformal gauge,
theory (\ref{1}) may be represented as some particular $D=2$
$\sigma$-model, hence it is expected that the effective action should
not depend on the choice of the fiducial metric as it usually happens in
string theory \cite{greenschwarzwitten87}. That is why the choice of
the reference metric in the form of the simplest --- flat space metric
--- has been made in ref.~\cite{russotseytlin92}.

However, not everything is clear in such a picture. In particular, there
are some subtleties in the definition of the path integral over metrics
in the $\sigma$-model approach to string theory. That may mean, that one
can find the explicit reference metric dependence in the calculation of
the off-shell effective action (at least in its finite part). Such
phenomenon maybe considered as some analog of the gauge
dependence of the off-shell convenient effective action in gauge
theories.

Motivated by these considerations, we study the off-shell one-loop
effective potential in the theory (\ref{1}) in the conformal gauge
\beq
g_{\mu\nu} =e^{2\sigma} \bar{g}_{\mu\nu},\label{4}
\eeq
where $\bar{g} _{\mu\nu}$ is not the flat space metric. In order to
estimate the influence of the topology on the effective potential as the
first example we choose $\bar{g} _{\mu\nu}$ corresponding to the metric
of topologically non-trivial plane $\cam =\reals \times S^1$. In the
second example $\bar{g}_{\mu\nu}$ corresponds to the metric of the
hyperbolic plane $\cam = H^2/\Gamma$ (so one can find the influence of
the topology and curvature on the effective potential). Note that for
hyperbolic manifolds, the co-compact discrete group $\Gamma$ in
$PSL(2,\reals
)$ is assumed to have only hyperbolic elements, so that the
resulting quotient manifold is a smooth one.

Hence, starting from the theory (\ref{1}) in the conformal gauge
(\ref{4}), and expanding near this static background, $\bar{g}_{\mu\nu}$
and $\bar{\varphi}$, where $\bar{\varphi}$ is the constant field, one
can easily obtain (see, for example, \cite{russotseytlin92})
\beq
\Gamma_{eff}^{(1)}[\bar{g}_{\mu\nu},\bar{\varphi}]=\frac 1 2 \ln {\det}
\Delta_{ij}-\ln\det(-\bar{\Delta}),\label{21} \eeq
where the second term in (\ref{21}) is the ghost contribution and
\cite{russotseytlin92}
\beq
\Delta_{ij}=\Delta_{ij}^{(0)}+M_{ij}\label{22}
\eeq
with
\beq
\Delta_{ij}^{(0)}=-\left(\begin{array}{cc}
                    \bar{\Delta} & 2c\bar{\Delta} \\
                     2c\bar{\Delta} & 0
                    \end{array}\right),\qquad
          M_{ij}=-  \left(\begin{array}{cc}
                  2V'' & 2V' \\
                           2V' & 4V
                          \end{array}\right).\label{23}
\eeq
Diagonalizing equation (\ref{21}) and dropping the non-essential
constants, one can find
\beq
\Gamma_{eff}^{(1)}[\bar{g}_{\mu\nu},\bar{\varphi}]=\frac 1 2 tr\left\{
\ln(-\bar{\Delta}+m_1^2)+\ln
(-\bar{\Delta}+m_2^2)-2\ln(-\bar{\Delta})\right\}\label{7}
\eeq
where
\beq
m_{1,2}^2=\frac 1
{2c^2}\left[V-2cV'\pm\sqrt{V^2-4cVV'+4c^2VV''}\right],\label{25}
\eeq
$V'=(\abl V/\abl\bar{\varphi})$ and
the trace is now only over a complete basis of the Hilbert space
$L^2(H^2/\Gamma)$ or $L^2(\reals\times S^1)$.
Note that we don't take into account the $V$-independent part of the
effective action, which gives the induced gravity term of the Polyakov
form, $\int R\Delta^{-1}R$.

It is seen, that equation
(\ref{7}) is equivalent to the knowledge of the determinant of the
operator
\beq D=-\Delta +M^2\label{8} \eeq
with some mass $M^2$.

For the hyperbolic plane under consideration very similar calculations
have already been done in $D=4$ dimensions
\cite{ser,bytsenkokirstenodintsov93,cognolakirstenzerbini93} and
for details of the calculation we refer
the interested reader to these references.
(Note that such a type of calculation of $Tr\ln$ on $D=3,4$ hyperbolic
manifolds is based on the use of the Selberg trace formula because the
explicit spectrum of the operator is not known. That is contrary to the
similar calculations on the sphere $S_4$
\cite{shore80,oconnorhushen83,allen83}, where the spectrum of the
operator is well-known.)
As a result, we may write the effective action in the form
\beq
\Gamma_{eff}^{(1)}[\bar{g}_{\mu\nu},\back ]
=B(m_1^2)+B(m_2^2)-2B(0).\label{9} \eeq
For the manifold $H^2/\Gamma$ after some work along the lines
of \cite{bytsenkokirstenodintsov93,cognolakirstenzerbini93}, for $M^2>0$
one may calculate $B(M^2)$ as ($\lambda$ is an arbitrary
renormalization scale) \beq B_{\Gamma}(M^2)&=&\frac{g-1} 2
\left[\delta^2+\frac 1
{12} \right]\ln\left(\frac{R_H^2\lambda^2}{\delta^2}\right)\label{10}\\
& &+\frac{g-1} 2 \delta^2+Y+T\nn
\eeq
with the abbreviations $\delta^2=M^2R_H^2+1/4$ ($R_H$ is the radius of
$H^2/\Gamma$), and \beq
Y=-(g-1)\ant_0^{\infty}dr\,\,r\ln\left(1+\frac{r^2}{\delta^2}\right)
(1-\tanh\pi r)\label{11}
\eeq
and
\beq
T=\frac 1 2 \ln \caz\left(\delta +\frac 1 2\right),\label{12}
\eeq
$\caz (s)$ being the Selberg zeta function corresponding to the group
$\Gamma$.

For $M=0$ the Laplacian has a zero mode. Thus in the limit $M\to 0$
equation (\ref{10}) has a logarithmic divergence, which is hidden in
$T$, equation (\ref{12}), due to $\caz (1) =0$. However, subracting as
usual the zero mode, leads to equation (\ref{10}) with $\delta^2=1/4$
and instead of equation (\ref{12}) use $T=(1/2)\ln\caz '(1)$.

For the manifold $\cam =\reals\times S^1$,
working along the lines of for example [14-18] (see [19] for a review),
\nocite{fordyoshimura79,ford80,toms80,denardospallucci80}
\nocite{denardospallucci80a,eli}
we have
\beq \frac{B(M^2)} R &=& -\frac{M}{2\pi}\sleu\frac 1 l
K_1(MLl)\label{13}\\
& & +\frac{\pi}{2L} \left(\frac{ML}{2\pi}\right)^2\left[1-
\ln\left(\frac M {\lambda}\right)^2\right]\nn
\eeq
useful for large values of $ML$, $L$ being the compactification length
of the torus, $R$ representing the (infinite) volume of the real line,
and \beq
\frac{B(M^2)} R &=&\frac 1 L \left[ {\pi}
\cq ^2\ln\left(\frac{L\lambda}{4\pi}\right)-\frac{\pi} 6 +\pi \cq +\pi
\gamma \cq ^2\right.\nn\\
& &\left.-\sqrt{\pi}\sum_{l=2}^{\infty}(-1)^l\frac{\Gamma\left(l-\frac 1
2\right)}{l!}\cq ^{2l}\zeta_R(2l-1)\right],\label{14}
\eeq
useful for small values of $ML$.

Now we can analyse the properties of our one-loop effective potential.
As one can see the dependence on the reference metric appears through
the non-zero effective masses in the effective potential. This
dependence may disappear for different types of potential. In
particular, for Liouville type potential,
\beq
V(\varphi )=\mu^2 e^{\alpha\varphi},\qquad \alpha =\frac 1
{2c}\label{15}
\eeq
with $\alpha$ defined as in (\ref{15}) (note that such a choice of $V$
corresponds to one-loop ultraviolet finite theory [4]) one can easily
see that $m_1=m_2=0$, the one-loop effective potential is equal to zero
and there is no reference metric dependence of the one-loop effective
potential. For other choices of the potential, we in fact have such an
dependence.

Now we may have a look to the effective equations corresponding to the
one-loop effective potential, $ \Gamma_{eff}=S +\Gamma^{(1)}$.
These equations are
\beq
\frac{\abl \Gamma_{eff}}{\abl \varphi}=0,\qquad
\frac{\abl \Gamma_{eff}}{\abl \rho}=0,\label{16}
\eeq
where $\rho =L$ for $\reals\times S^1$ space and $\rho =R_H$ for the
hyperbolic plane. Note that if we would want to use the Kaluza-Klein
type interpretation we could put by hands the additional
condition---vanishing of the effective cosmological constant---
$\Gamma_{eff}=0$ on the solutions of equations (\ref{16}).

The effective equations are of course easily found from the expressions
we derived for $\Gamma_{eff}^{(1)}$. To give solutions of these
equations we consider for simplicity the Liouville type potential,
equation (\ref{2}). Thus
\beq
m_1^2=c^{-2}\mu^2e^{\alpha \back}(1-2\alpha c),\qquad m_2^2=0.\label{19}
\eeq
Combining equations (\ref{16}) in a useful way, for $H^2/\Gamma$ we find
\beq
m_1^2R_H^2=\frac 1 3 , \qquad \delta^2 =\frac 7 {12}.\label{20}
\eeq
Thus the on-shell effective action for this case is simply given by
(\ref{9}), (\ref{10}), with $\delta^2=7/12$.

For $\reals\times S^1$ the effective equations are not so simple,
however using these effective equations it is possible to write the
on-shell effective action in the form \beq
\frac{\Gamma_{eff}} R =-\frac 1 L \left\{L^2V(\back)
-\frac{(Lm_1)^2}{4\pi}\left[
1-\ln\left(\frac {m_1} {\lambda}\right)^2\right]-\frac{\pi}
3\right\}\label{30}
\eeq
Thus, from our two explicit examples one can see, that not only the
off-shell one-loop effective potential in $2D$ dilaton gravity in
conformal gauge, but also the on-shell one depends explicitly on the
reference metric. If we interpret this phenomenon as some kind of the
gauge dependence of the convenient effective potential in gauge theories
(like scalar QED), we see perfect analogy between them. Indeed, in gauge
theories the off-shell effective potential is usually gauge dependent,
and the on-shell effective potential is gauge independent in a sense,
that a general theorem exists for the formal non-perturbative on-shell
effective potential. However, in loop expansion the explicit calculation
shows the gauge dependence of the on-shell one-loop effective
potential. Extending this analogy further, one would expect that the
non-perturbative on-shell effective potential in $2D$ gravity should
not depend on the reference metric. It would be of interest to analyse
this question further.

\section*{Acknowledgments}
KK and SDO would like to thank the members of the Department ECM,
Barcelona University, for the kind hospitality.
SDO would like to thank A.~Tseytlin for helpful remarks and CIRIT
(Generalitat de Catalunya) for financial support.
KK acknowledges helpful discussions
with Sergio Zerbini and
furthermore financial support from the Alexander von
Humboldt Foundation
(Germany) and from CIRIT (Generalitat de
Catalunya).

\end{document}